\documentclass[titlepage,12pt]{article}
\usepackage{amsmath,amssymb,amscd,graphicx,xy,epsfig}

\xyoption{matrix}\xyoption{arrow}\xyoption{frame}

\newcommand{\be}{\begin{eqnarray}}
\newcommand{\ee}{\end{eqnarray}}
\begin{document}
\addtolength{\baselineskip}{1mm}
\setlength{\parskip}{.5mm}

\addtolength{\abovedisplayskip}{1.1mm}
\addtolength{\belowdisplayskip}{1.1mm}
\newcommand{\is}{ \!&\! = \!& \!}
\newcommand{\overgs}{\raisebox{-.5mm}{\large ${1\over 2g_s}$} }

\renewcommand{\footnotesize}{\small}

\begin{titlepage}
\begin{center}

{\hbox to\hsize{
\hfill PUPT-2215}}

{\hbox to\hsize{\hfill hep-th/0611069}}

\vspace{3cm}

{\Large \bf 
On Metastable Branes} \\[5mm]
{\Large \bf and a } \\[5mm]
{\Large \bf 
New Type of 
Magnetic Monopole}\\[2cm]
{\large Herman Verlinde }\\[10mm]

{\it Department of Physics \\[2mm] Princeton University\\[2mm] Princeton,
NJ 08544, USA}\\[4mm]


\vspace*{2.5cm}

\end{center}
\noindent
String compactifications with D-branes may exhibit
regular magnetic monopole solutions, whose presence does not rely on 
broken~non-abelian~gauge~symmetry.~These~stringy~monopoles exist
on interesting metastable brane configurations, such as anti-D3 branes 
inside a flux compactification or D5-branes wrapping 2-cycles that are 
locally stable but globally trivial.
In brane realizations of SM-like gauge theories, the monopoles carry one unit of 
magnetic hypercharge. Their mass can range from the string
scale down to the multi-TeV regime.

\end{titlepage}
\newpage
\tableofcontents

\renewcommand{\footnotesize}{\small}
\newpage

\newcommand{\ba}{\begin{eqnarray}
\addtolength{\abovedisplayskip}{1.1mm}
\addtolength{\belowdisplayskip}{1.1mm}}
\newcommand{\ea}{\end{eqnarray}}
\newcommand{\bea}{\begin{eqnarray*}}
\newcommand{\eea}{\end{eqnarray*}}

\newcommand{\FF}{{\mathsf F}}
\newcommand{\sss}{{\! s}}
\newcommand{\aaa}{{\mbox{\scriptsize \sc a}}}
\newcommand{\bbb}{{\mbox{\scriptsize \sc b}}}
\newcommand{\N}{{\mathcal{N}}}
\newcommand{\D}{{\mathcal{D}}}
\newcommand{\La}{{\mathcal{L}}}
\newcommand{\OO}{{\mathcal{O}}}

\renewcommand{\SS}{{\bf S}}
\newcommand{\BB}{{\bf B}}
\newcommand{\Z}{{\mathbb Z}}

\newcommand{\ub}{\underbrace}

\newcommand{\Appendix}[1]{
\addtocounter{section}{1} 
\setcounter{subsection}{0} \addcontentsline{toc}{section}{\protect
\numberline{\arabic{section}}{{\rm #1}}} \vglue .0cm \pagebreak[3]
\noindent{\large \bf  \thesection. #1}\nopagebreak[4]\par\vskip .3cm}

\newcommand{\newsection}[1]{
\addtocounter{section}{1} 
\setcounter{subsection}{0} \addcontentsline{toc}{section}{\protect
\numberline{\arabic{section}}{{\rm #1}}} \vglue .0cm \pagebreak[3]
\noindent{\large \bf  \thesection. #1}\nopagebreak[4]\par\vskip .3cm}
\newcommand{\newsubsection}[1]{
\addtocounter{subsection}{1}
\addcontentsline{toc}{subsection}{\protect
\numberline{\arabic{section}.\arabic{subsection}}{ #1}} \vglue .0cm
\pagebreak[3] \noindent{\bf \thesubsection. #1}\nopagebreak[4]\par\vskip .3cm}

\newsection{Introduction}

In this note we point out that string compactifications with D-branes 
naturally contain
an interesting yet unfamiliar type of magnetic monopole excitations. These monopoles are smooth finite
energy solutions, and appear in many string models with abelian gauge
symmetries associated with D-branes. 
Their regularity and stability does not depend on the presence of broken 
non-abelian gauge symmetry. 
The monopoles arise in two rather similar situations,
both of which involve interesting types of metastable D-branes.

Both in string theory and in gauge theory, it has
become clear that vacua with broken supersymmetry are typically 
metastable against decay via vacuum tunneling \cite{KKLT,KPV,DDGR,ISS}. 
Metastable D-brane configurations are of direct physical
interest, as examples of non-supersymmetric theories with controllable
dynamics \cite{KPV,susybog,metabranes}, or as an accessable corner within the non-supersymmetric string landscape \cite{flux}. In our set-up, the 
metastability is induced by the geometry of the compact cycle wrapped by the branes. 

Specifically, we will consider D5-branes  wrapped on a metastable 2-cycle $\alpha$ 
within a compact Calabi-Yau geometry. The D5-branes are supported either by magnetic flux 
on the brane or by the local CY geometry. Globally within the Calabi-Yau manifold, however, 
the 2-cycle $\alpha$ is trivial: there exists a 3-chain $\Gamma$ which has the 2-cycle $\alpha$
as its boundary. 
This means that, topologically, the D5-branes can self-annihilate by unwrapping around $\Gamma$. 
In our set-up, we will assume that the 2-cycle $\alpha$ as small compared to the 3-chain
$\Gamma$, so that in the unwrapping process,  the branes must first expand 
and thereby increase their energy. The branes are therefore classically stable, but
have a finite lifetime against decay by quantum tunneling through the potential barrier. 
The decay rate can be made exponentially small  by ensuring that, while unwrapping, the 
D-branes must traverse a large enough 3-cycle.

The appearance of regular magnetic monopole excitations is directly linked to
the metastability of the D5-branes. Namely, consider a D3 and D5-brane that both wrap
the same metastable 2-cycle. Assuming the D3-brane is classically stable, it 
represents a Nielsen-Olesen flux tube within the D5 world volume. By design,
the D3-brane is metastable and can unwrap via tunneling. The unwrapping process in
interpreted in the gauge theory as the breaking of the magnetic flux tube via 
the formation of a monopole anti-monopole pair. The monopole is a D3-brane
wrapping the 3-chain $\Gamma$.

This new class of stringy monopoles are generic in string scenarios with 
D-branes at CY singularities, and are typically associated with massless $U(1)$ 
gauge factors. In D-brane realizations of SM-like gauge theories, 
the monopoles carry magnetic hypercharge. In highly warped scenarios with low (local) 
string scale, the monopole mass can be less than 100 TeV.

\bigskip

\bigskip

\bigskip


\newsection{Metastable D5-branes}

In this section we introduce two examples of geometrically metastable branes:
anti-D3-branes in a flux compactification, and D5-branes wrapping trivial 2-cycles.

\bigskip
\bigskip

\newsubsection{Geometric Setting}

\noindent
{\em Setting I:  Anti-D3 in a  IIB Flux Compactification}

\smallskip
 
The first type of situation has a very well-known and familiar example: a single 
anti-D3 brane in a IIB flux compactification,
as 
used in the KKLT construction. For our purpose, the relevant property of this set-up
is that the anti-D3 brane is 
located at the intersection of  two 3-cycles $A$ and $B$ with non-zero 3-form fluxes  
\ba
\label{een}
\int_A F_{\it 3} = M\, , \qquad \qquad \int_{B} H_{\it 3}\, =\,  K. 
\ea 
In a supersymmetric compactification geometry, the RR 3-form field strength $F_3$ and NS 
3-form field strength $H_3$ combine into a single imaginary anti-self dual field strength
$G_3$. The presence of the anti-D3 brane breaks the supersymmetry. For
an anti-D3 brane localized at the tip of a conifold, it was shown in \cite{KPV} that
this situation is classically stable but quantum mechanically metastable: the anti-D3 brane 
can decay via brane-flux annihilation.

To visualize the decay
process, one needs to think of the anti-D3 brane as a small spherical 5-brane, supported by one 
unit of magnetic flux through the $\SS^2$ representing the compact part of its world-volume:
\be
\label{above}
\int_{S^2} F_{\it 2} = -1\, .
\ee
The magnetic flux induces an anti-D3 charge, and protects the 5-brane from  
self-annihilation. Instead it is trapped in a metastable non-supersymmetric state. The 5-brane 
can decay into supersymmetric final state by unwrapping itself, by traversing the A-cycle or 
the B-cycle.  The decay is suppressed by a potential barrier, since in the unwrapping process, 
the 5-brane first has to expand.  Which of the two options (traversing the A-cycle or B-cycle) is 
most likely,  thus depends on the relative size of the two cycles.  
The decay along the A-cycle has been analyzed in detail in \cite{KPV}. In section 2.3,
we will summarize the decay mechanism along the B-cycle, which follows via a simple translation of the 
analysis in \cite{KPV}. 
 Both decay rates are assumed to be very slow.

\bigskip

\bigskip

\noindent
{\em Setting II: D5-brane Wrapping a Trivial Cycle}


Consider a Calabi-Yau threefold $Y$ with a cone-like singularity. The base $X$ of the singularity 
is 2-d complex submanifold within $Y$ and in general contains several 2-cycles.  
All these 2-cycles are non-trivial within $X$, but some may be trivial within the threefold $Y$.  
Such a trivial 2-cycle $\alpha$ is  the 
boundary of a 3-chain $\Gamma$ within $Y$
\be
\label{drie}
\partial \Gamma = \alpha\, .
\ee 
The 2-cycle $\alpha$ is located at the tip of the cone and assumed
to be very small. The 3-chain $\Gamma$, on the other hand, typically extends 
into the bulk of the CY threefold and can be comparatively large. A type II
string compactification on $Y$ may thus contain D-branes, that wrap $\alpha$.
These D-branes are classically stable, but topologically and quantum mechanically, 
they can self-annihilate by unwrapping around the 3-chain $\Gamma$.  
Since the 5-brane first has to expand in the unwrapping process, the decay rate 
may be highly suppressed.

This class of geometrically metastable D5-branes are of special interest for a separate reason:
unlike D-branes wrapped on non-trivial 2-cycles, they can support a massless
$U(1)$ vector boson. The 3+1-d world-volume theory of D-branes on a compact CY threefold 
generally features a the $C\! \wedge\! F$ coupling  to an RR 2-form $C$. In our case,
this 2-form $C$  equals~the integral of the RR 4-form  potential over the 2-cycle wrapped
by the D5-brane. If this 2-cycle is non-trivial within $Y$, the 2-form $C$ has a normalizable 
kinetic energy. Combined with the $C\! \wedge\! F$ interaction, this produces a mass
term for the $U(1)$ vector boson. This is the~familiar stringy St\"uckelberg mechanism. 
If, on the other hand, the D5-brane wraps a trivial
2-cycle $\alpha$, then 
\ba
\int_\alpha \omega = 0
\ea
for every normalizable harmonic 2-form $\omega$ on $Y$. This means that the St\"uckelberg field
is absent, and that the abelian gauge boson on the brane world-volume remains massless.  (This mechanism was recently used in \cite{BMMVW} to obtain a D-brane realization of an SM-like
gauge theory with only hypercharge as massless abelian gauge generator.)

The D5-brane may in general wrap $p$ times around the 2-cycle $\alpha$. In addition,
it can support some units of magnetic flux 
\ba
\label{qflux}
\int_\alpha F_{\it 2} = q.
\ea
In brane language, this means that the $p$ D5-branes carry $q$ units of D3-brane charge. 
We will call this
D5/D3 bound state a ``fractional brane" with charge vector $(p,q)$. 
If $q$ and $p$ are both positive, then the fractional brane is typically supersymmetric. 

If $p$ and $q$ are mutually prime, the 3+1-d world volume theory of the
fractional brane has $U(1)$ gauge symmetry.\footnote{The 5+1-d 
$U(p)$ gauge symmetry of the D5-brane is broken to $U(1)$ by the magnetic flux~(\ref{qflux}). For a more detailed summary of the properties of fractional branes at CY singularities, see \cite{MH,BMMVW} and, mostly, the references therein.}
The gauge coupling is given by the general formula
\ba
{4\pi \over e^2} = {1\over g_s} \, |{\mathsf Z}_{p,q} |
\ea
where ${\mathsf Z}_{p,q}$ denotes the central charge of the fractional brane. The central charge is 
complex number that tells us which half of the supercharges is preserved by the CFT boundary state of the fractional
brane. It is a linear function of the charge vector
\be
\label{zee}
{\mathsf Z}_{p,q} =  q + p\, (b + i v)\, .
\ee
In the large volume regime, where the 2-cycle $\alpha$ is large compared to the string
scale, we can make the geometric identification
\ba
b =\!  \int_\alpha B_{\it 2} \, , \qquad \qquad \ \ v = \! \int_\alpha J\, ,
\label{eight}
\ea
where $B_{\it 2}$ is the NS 2-form, and $J$ denotes the K\"ahler 2-form on $Y$. 
Using the invariance of the IIB string theory  under integral shifts 
\be
b \to b - n\, ,  \qquad \qquad \ 
q \to q + n p\, . 
\ee
we will define the B-field period $b$ as an angular variable between 0 and 1. 

The 3-chain $\Gamma$ may in general support a non-zero NS 3-form flux equal to
\be
\label{hflux}
\int_\Gamma H_{\it 3} 
=  b + K
\ee
with $K$ some integer. This flux will have important physical consequences in the following.

For finite K\"ahler area $v$ of the trivial
2-cycle $\alpha$, the world volume gauge theory of the fractional brane has a non-zero D-term.
In case there are other fractional branes at the singularity, this D-term turns into an FI-parameter
that dictates non-zero expectation values for charged matter fields. This breaks the $U(1)$
gauge symmetry and again renders the abelian gauge boson massive. In the following, 
we will therefore mostly consider the case where $\alpha$ has  zero volume,  $v=0$,
so that the gauge multiplet  stays massless. In this case, the regular monopole solutions, that
we will exhibit, are unconfined and a smooth source of a rotationally symmetric 
magnetic Coulomb field.

\bigskip

\bigskip

\bigskip

\newsubsection{World-volume Action}

The two settings described above are very similar. In a rather concrete sense, one can view
Setting I as a special case of Setting II, in which (i) the 2-cycle $\alpha$ is a 2-sphere,
(ii) the B-field period vanishes, $b=0$,
and (iii)  the fractional brane is just a spherical D5-brane with one unit of anti-D3 charge,
with charge vector $(p,q)=$ {\small (1,$-$1)}. We will now describe the world-volume 
DBI action of the geometrically metastable D5-brane of Setting II. For simplicity, we will
set the D5-brane wrapping number equal to $p=1$.

We focus our attention to the space-time  ${\mathbb R}^{3,1} \times \Gamma$ traversed by the 
D5-brane during the unwrapping process along  $\Gamma$. 
We write the metric 
as
\ba
ds^2  \is  \, dx_\mu dx^\mu\, + 
\, d{\phi}^2 \; 
+  \; A(\phi) \, d s_2^2 
\label{induced}
\ea
Here we ignored any warp factors.
The coordinate $\phi$ parametrizes a slicing of 3-chain $\Gamma$. 
The metric $ds_2^2$ is normalized 
so that the function $A(\phi)$ represents the physical area of the cross section at the location $\phi$.
The coordinate range of $\phi$ is finite, from 0 to some maximal value $\phi_*$. 
We will denote the cross-section at location $\phi$ by $\alpha(\phi)$. The initial cross-section 
$\alpha(0)$ is the 2-cycle $\alpha$; the final cross-section $\alpha({\phi_*})$ is a point.

\newcommand{\sgs}{{\raisebox{-.5mm}{\small 1} \over\raisebox{.5mm}{ $g_s$}}}

\newcommand{\sqgs}{{\raisebox{0mm}{\small $|Q|$} \over\raisebox{.5mm}{ $g_s$}}}

\newcommand{\sqphi}{{\raisebox{0mm}{\small $Q^2$} \over\raisebox{0mm}{ $4\pi \phi^2$}}}

\smallskip

Consider a D5-brane at some transverse location $\phi$. 
This location can fluctuate and $\phi$ thus  represents a scalar field on 
the D5 world-volume. The bosonic D5-brane action reads 
\ba
S = \frac{1}{g_s}
\int\! \sqrt{\mbox{\small $\det{(G_\parallel + F)}
\det(G_\perp \! + {\cal F})$}}\,
\; +  \int  C_{\it 6}\, ,
  \\[4.5mm]
F =\, F_{\mu \nu}\, , \qquad \qquad 
{\cal F}_{\it 2}\!  = F_{\it 2} - B_{\it 2}\, .\qquad 
\ea
Here $F_{\mu\nu}$ and $F_{\it 2}$ are the non-compact and compact components of 
the field strength of the world-volume gauge field, $G_\perp$ denotes the induced 
metric along the internal
$\SS^2$, and $G_\parallel$ encodes the remaining components along the $\phi$ and
${\mathbb R}^4$ directions. Evaluating the various factors gives\footnote{Here we use 
that, in the supersymmetric IIB background, the total 3-form flux $G_{\it 3} = F_{\it 3} -\tau H_{\it 3}$
is imaginary anti-self-dual. }
\ba
\label{BI}
S \is {1\over g_s} 
 \int \Bigl(V(\phi) \sqrt{\det\bigl(\eta_{\mu\nu}  +   F_{\mu\nu}  + \partial_\mu \phi\, \partial_\nu\phi\bigr)} - B(\phi)\Bigr)
\\[5mm]
& & 
\ \qquad V(\phi) = \sqrt{(A(\phi))^2 + (B(\phi))^2}
\ea
where $B(\phi)$ is defined via
\begin{equation}
\label{twee}
 \int_{\alpha(\phi)} \! \! {\cal F}_{\it 2} =  B(\phi)\,.
\end{equation}

Let us give a qualitative description of  the functions $A(\phi)$, $B(\phi)$ and $V(\phi)$ that appear 
in this action. The boundary values of these three functions are 
\ba
\label{bb}
A(0) = v \ \ \ \ \ \qquad \qquad \quad \ &  & \qquad A(\phi_*) = 0 \nonumber  \\[2.6mm]
B(0) = q + b \, \qquad \qquad \quad\; & & \qquad B(\phi_*) =  q - K \\[3mm]
V(0) = \sqrt{ v^2 + (q+b)^2}    \ \ \ & & \qquad V(\phi_*) = |q- K| \nonumber
\ea
Here $v$ is the area and $b$ the B-period of the 2-cycle $\alpha$ (see eqn (\ref{eight})). Note that
$V(0) = |Z_{1,q}|$ with $Z_{1,q}$ the central charge (\ref{zee})
of the metastable brane configuration. 

The generic profile of the three functions is as given in fig 1. $A(\phi)$ starts at some very small
value $v$, grows to a maximal value, then decreases until it reaches 0 at $\phi=\phi_*$. $B(\phi)$ is a 
monotonic function. The total potential energy of the D5-brane is
\ba
\label{vtot}
V_{\rm tot}(\phi) =  {\raisebox{-.5mm}{\small 1} \over\raisebox{.5mm}{ $g_s$}}\bigl(\, V(\phi) - B(\phi)\bigr).
\ea
Its initial and final values are
\ba
V_{\rm tot}(0) \is  {\raisebox{-.5mm}{\small 1} \over\raisebox{.5mm}{ $g_s$}} \bigl(\sqrt{v^2 + (q+b)^2}\, - q- b\Bigr)\nonumber \\[3mm]
 V_{\rm tot}(\phi_*) \is  {\raisebox{-.5mm}{\small 1} \over\raisebox{.5mm}{ $g_s$}}\bigl(\, |q-K| - q+K\bigr)
 \ea
Assuming that $q\!-\!K\!>\!0$, this total potential energy vanishes at
$\phi\!=\!\phi_*$. This indicates that~the~situation after complete 
unwrapping of the D5-brane is supersymmetric.  The 5-brane 
does not completely dissolve at this point, however: its tension remains finite and
equal  to $q- K$ times that of a D3-brane. The final situation indeed contains $q - K$
D3-branes, that are all supersymmetric relative to the flux background. In the special case that
$q=K$, the unwrapped state has no D3-branes. The BI gauge theory (\ref{BI}) then becomes infinitely
strongly coupled at $\phi=\phi_*$. Indeed all open string degrees of freedom are confined at that point,
and the D5-brane has dissolved into pure closed strings.


\begin{figure}[t]
\begin{center}
\leavevmode\hbox{\epsfxsize=11cm \epsffile{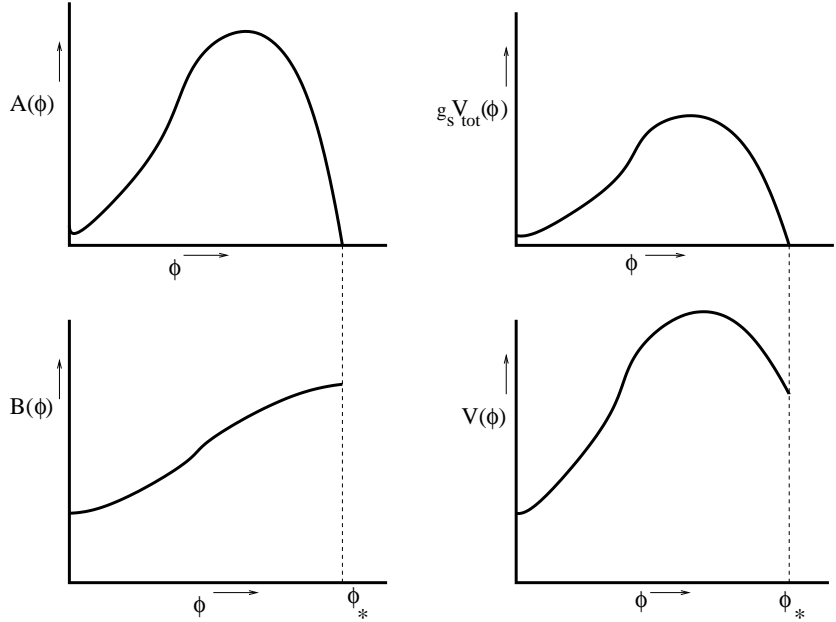}}
\caption{The typical profile of the area $A(\phi)$, effective D3-charge $B(\phi)$,
tension $V(\phi)$, and total potential energy $V_{tot}(\phi)$ of the D5-brane as a function of the
transverse location~$\phi$. }
\end{center}
\end{figure}

In the metastable situation, at $\phi=0$, the potential energy is generally positive, but 
vanishes in the limit $v\to 0$, provided that $q>0$.  In this limit, the trivial 2-cycle
$\alpha$ has zero size. 
In Setting I, we have $b=v=0$, and $q=-1$. 
The total potential energy of the D5-brane at $\phi=0$ then
equals twice the D3 brane tension. This
is the expected result for a single anti-D3 brane in an otherwise supersymmetric 
background \cite{MN,KPV}. (The factor of two arises, since an anti-D3 brane can be thought
of as a linear superposition of a supersymmetric ghost brane with opposite charge and tension 
of an D3-brane and a chargeless brane with twice the D3-tension.) 
Similarly, for $q - K\!< 0$  the potential energy at $\phi = \phi_*$ equals twice 
the~tension of $q-K$ anti-D3 branes.

The function $B(\phi)$ is the interpolating function that keeps
track of the D3-charge carried by the wrapped D5-brane. In traversing the 3-chain
$\Gamma$, the D5-brane thus acquires a total of $K\!-\! b$ units of D3-charge. 
The total D3-charge is still conserved, however: the change in the charge carried by the D5-brane 
is compensated via a jump by one unit in the RR-flux 
through the 3-cycle dual to~$\Gamma$. (Recall that the 3-form flux carries a D3-charge density
proportional to $F_{\it 3} \wedge H_{\it 3}$, so that if the dual~$F_3$~flux jumps
by one unit, the integrated charge changes by an amount equal to  the total $H_{\it 3}$-flux through~$\Gamma$.)

\bigskip
\bigskip
\bigskip

\newsubsection{Vacuum Decay}

A semi-quantitative description of the decay of the metastable
D5-brane configuration is obtained as follows. 
In space-time, the decay takes place via 
nucleation of a bubble of supersymmetric vacuum surrounded by a
spherical domain wall which expands exponentially
as a consequence of the pressure produced by the drop in the vacuum
energy. To obtain the nucleation rate, it is standard practice to look for a
corresponding Euclidean solution. The relevant solution for us
is a D5 brane trajectory $\phi(R)$, where $R$ is the radial
coordinate in ${\mathbb R}^4$, connecting the wrapped D5-brane at $\phi=0$ 
at large $R$ to an instantonic domain wall located near some appropriate radius
$R = R_*$, beyond which the solution reaches the supersymmetric minimum
$\phi = \phi_*$.
The action functional for such a trajectory reads
\be
S = 
{2\pi^2\over g_s}
\int_{R_*}^\infty \!\! dR\, R^3 \, \Bigl( \, {V}(\phi)
\sqrt{ 1  +  (\partial_R{\phi})^2}
+ B(\phi) \Bigr)
\ee

One can now obtain a semi-classical estimate of the nucleation rate by evaluating the 
action of the (numerically obtained) bounce trajectory.  
In case the potential barrier is much higher than the energy difference between the
initial and final configurations, this calculation can be simplified via the familiar 
``thin wall approximation'' \cite{Coleman}.  In this approximation, the action reduces to
the sum of two terms, representing  the two competing forces on the 
5-brane domain wall bubble: the tension pulling it inward and the outward pressure 
caused by the lower energy of the supersymmetric vacuum:
\be
S\; = \, 2\pi^2 R_*^3 \; {\cal T} \,  - \;
\textstyle {1\over 2} \pi^2 R_*^4 \; \Lambda \, .
\ee
Here ${\cal T}$ is the domain wall tension, $\Lambda$ the drop in vacuum energy,
and $R_*$ denotes the location of the wall. At the critical radius
\be
\label{rcrit}
R_* = {3 {\cal T}/\Lambda}
\ee
the two forces are balanced. Plugging $R_*$ back into the action, 
gives the leading estimate of the nucleation rate
\ba	
\label{finalrate}
\Gamma \simeq\, \exp\big(\, {\small \!-\!} {\mbox{ \large $ {27 \pi^2 \over 2}$}\! \mbox{ \large $
  {{\cal T}^4 \over \Lambda^3 }$}}\, \bigr) 
\ea
In our problem, the wall tension and energy difference between the two vacua are given by 
\ba
\label{tension}
\qquad {\cal T} \is {\raisebox{-.5mm}{\small 1} \over\raisebox{.5mm}{ $g_s$}}\, V_B 
\qquad \qquad \quad \ 
V_B =\! \int_0^{\phi_*}\!\!\!\! d\phi\, A(\phi) \\[3mm]
\Lambda \is  {\raisebox{-.5mm}{\small 1}\over \raisebox{.5mm}{\small $g_s$}}
\bigl(V(0) - B(0)\bigr) \, = \, V_{\rm tot}(0)
\ea
The quantity $V_B$ equals the 3-volume of the B-cycle. 

The result (\ref{tension}) for the domain wall tension
is obtained as follows. 
Consider the limit where $\Lambda = 0$.  
The initial and final configurations with $\phi=0$ and $\phi=\phi_*$ are then both supersymmetric, and
the tunneling rate vanishes. There is however still a BPS domain
wall that interpolates between the two vacua. Let us find its tension. The domain wall is   
described by D5-brane following
 a spatial trajectory $\phi(z)$  with $z$ the coordinate
transverse to the wall.
The $z$-evolution of $\phi(z)$ is governed by the Hamilton equations $\partial_z\phi = {2\pi^2\over g_s} {\partial {\cal H} \over \partial P_\phi}$ , $\partial_z P_\phi = {2\pi^2\over g_s} {\partial {\cal H} \over \partial \phi}$ generated by 
\ba
{\cal H} =
\sqrt{\bigl( \, 
A(\phi)^2 + B(\phi)^2 - P_\phi^2\bigr)} \, - \, 
B(\phi) \, .
\ea
The BPS solution has ${\cal H}=0$, which gives $P_\phi\! =\! A(\phi)$.
The Hamilton equation then reads\footnote{Note that for $q >K$, the function $B(\phi)$ remains positive throughout the interval from 
0 to $\phi_*$. This is not the case for the non-supersymmetric situation $q<K$.}
\ba
\partial_z \phi = {A(\phi)\over  B(\phi)}\, .
\ea
The result (\ref{tension}) follows by evaluating the action of this BPS solution. The BPS formula is
appropriate for use in eqn (\ref{finalrate}) , since in the thin wall approximation one assumes that the decay takes place  between two nearly degenerate vacua.

\bigskip

\bigskip
\bigskip

\bigskip

\bigskip

\newsection{Stringy Monopoles}

Consider the bound state of a D5-brane and a D3-brane wrapping a trivial 2-cycle $\alpha$ inside a compact Calabi-Yau. Assume the 2-cycle has a small but finite area.
This configuration represents a Nielsen-Olesen flux tube within the 
$U(1)$ gauge theory on the D5-brane world-volume. 
The tension of the magnetic flux tube is proportional to the $U(1)$ symmetry
breaking scale, which in the gauge theory/geometry dictionary is set by the area of the 2-cycle. 
It is natural to ask whether the flux tube is stable or  can break via the formation of a 
monopole anti-monopole pair. 

In light of the above discussion of the brane decay, the 
answer is obvious:  the magnetic flux tube can break because 
the D3-brane can unwrap itself by traversing
the 3-chain $\Gamma$. The monopole at the end flux tube is  thus identified as a D3-brane 
that wraps $\Gamma$.  This interpretation nicely matches with standard identification for end-points 
of D(p$-$2)-branes on Dp-branes. In most studies so far, however, the 
D(p$-$2)-brane would either be an semi-infinite BI-on of infinite mass \cite{bion}, or would end 
on an other Dp-brane and thus represent a regular non-abelian monopole of finite mass, proportional
to the distance between the two Dp-branes \cite{diaconescu}. 
Our situation is different, and novel: the open D3-brane, that makes up the monopole, extends over a finite range, yet does not end on
any other brane.\footnote{Monopoles arising from D3-branes wrapping 3-chains were first considered in \cite{braneonchain},
in a situation without space-time filling world-branes. The abelian gauge theory then
arises from the RR 4-form field $C_{\it 4}$. The monopoles are necessarily confined since the
wrapped D3-branes can not end, but must be connected by a D3-brane wrapping the trivial 
2-cycle $\alpha$.}

\bigskip

\bigskip

\newsubsection{Monopole Configuration}

Our monopoles are confined via a magnetic flux tube as long as
the 2-cycle $\alpha$ has finite size. When the 2-cycle has zero size, the 
abelian gauge bosons on the D5-brane become massless and the monopoles are liberated. 
The D3-brane charge then gets dissolved into the flux lines of the 
magnetic Coulomb field on the D5-brane world-volume.
The resulting D-brane configuration is described as follows.

Consider a D5-brane world volume, obtained by gluing together the following 
two parts. The first part is a D5-brane that wraps the 2-cycle $\alpha$ and
fills $\mathbb R^3$ minus a three-ball $B_3$. 
The world volume of this part of the D5-brane (we ignore the time direction) is
\ba
\Sigma_1 = ({\mathbb R}^3 - \BB_3) \times \alpha
\ea
The three-ball $\BB_3$ has as boundary a 2-sphere $\SS^2$. 
The second part of the D5-brane worldvolume
wraps this $\SS^2$ and the 3-chain $\Gamma$:
\ba
\Sigma_2 = \SS^2 \times \Gamma\,.
\ea
Using eqn (\ref{drie}) and $\partial \BB_3 = \SS^2$, we see that the two halves $\Sigma_1$ and $\Sigma_2$ have as common boundary 
\ba
\partial \Sigma_2  = \SS^2 \times \alpha =  - \partial \Sigma_1\, .
\ea
The two halves   can thus be 
glued together into a single, smooth D5-brane world-volume.  

When left on its own, this D5-brane configuration would be unstable against collapse of  the
$\SS^2$ to a point. However, we can now introduce a quantized magnetic flux 
\be
\int_{S^2} F = Q
\ee
through the 2-sphere. The energy contained in this magnetic flux stabilizes the brane, by forcing the 2-sphere to remain at some finite size. Its radius $r$ depends on the transverse location $\phi$  
of the D5-brane along $\Gamma$. In other words, the D5-brane solution is
specified by its $\phi$ location as a function of radial coordinate $r$. The precise form of
the trajectory $\phi(r)$ is determined by minimizing the energy functional
\ba
\label{mmass}
M = {1\over g_s} \int \! dr\, \Bigl(  V(\phi) \sqrt{ 
(16\pi^2 r^4 + Q^2) \, 
\bigl(1+ {(\partial_r\phi)^2}\bigr)}- 4\pi r^2 B(\phi) \Bigr)\, .
\ea
The total brane configuration is depicted in fig 2. Evidently,
it represents a smooth, spherically symmetric finite energy excitation with
 the physical properties of an abelian magnetic monopole of charge $Q$. Note that
 there is symmetry between monopoles and anti-monopoles.

The D5-brane has the shape of a finite section of the Bion funnel. In 
the {$\Sigma_2$~region~it} looks like
a puffed up D3-brane wrapping the 3-chain $\Gamma$. We thus expect the mass of the~monopole to be given by the volume of $\Gamma$, in units of the D3-tension, plus some self-energy correction:
\ba
M = 
{|Q|\over g_s} \int_0^{\phi_*}\! \! \! d\phi\, V(\phi)\; \, + \, \;  E_{\rm self}
\,
\ea
One can derive upper bounds for $E_{\rm self}$, by evaluating (\ref{mmass}) for the Bion trajectory $\phi = {|Q| / 4\pi r}$ and other simple trajectories. There is no obvious exact formula: the monopole is a non-BPS solution.

\begin{figure}[t]
\begin{center}
\leavevmode\hbox{\epsfxsize=7cm \epsffile{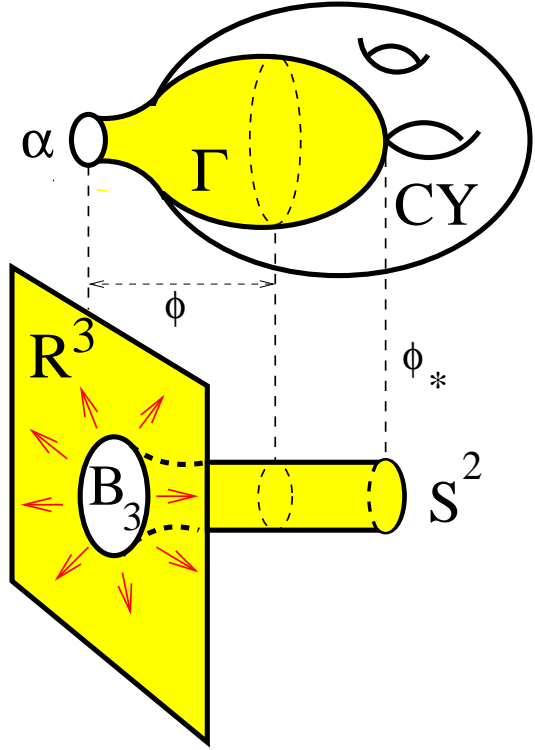}}
\caption{The D5-brane configuration that describes the smooth monopole solution. The D5
world brane wraps the trivial 2-cycle $\alpha$, but near the origin unwraps via the
3-chain $\Gamma$.}
\end{center}
\end{figure}

This all looks quite consistent, but there is one subtlety that we have ignored so far:
as with the D5-brane decay,  the presence of $H_3$ flux (\ref{hflux}) through the 3-chain $\Gamma$ 
will lead to a non-trivial induced D1-charge on the D3-brane world-volume. So to
ensure that the above D5 brane solution really rounds off smoothly at
the tip $\phi=\phi_*$, we need to require that it supports no ${\cal F}_2$ flux through
$\alpha(\phi_*)$. This amounts to setting (see eqns~(\ref{twee})-(\ref{bb}))
\be
B(\phi_*) = q - K = 0\,.
\ee
In other words, the monopoles exist as smooth finite energy solutions on metastable
D5-branes with $K$ units of D3-charge.

\bigskip
\bigskip

\newcommand{\sumk}{\mbox{\large $\sum\limits_{\raisebox{.5mm}{\scriptsize $k$}}$}}

\newcommand{\qq}{{\mathsf q}}
\newcommand{\pp}{{\mathsf p}}
\newcommand{\rr}{{\bf  r}}

\newsubsection{Hypercharge Monopoles}

The new monopole excitation also makes its appearance in open string realizations of 
Standard Model gauge theories, via D-branes at CY singularities. In this context,
the physical role of the metastable D5-brane is slightly more subtle, since it in fact 
does  not represent one of the fractional branes, but rather a linear combination of
fractional branes. Let us briefly explain this basic  point in the general context of D-branes at
singularities.

A basis of fractional branes associated with a CY singularity spans the complete homology of 
its base $X$. 
Let us assume that $X$ forms 
a 4-cycle within $Y$, and supports several 2-cycles 
 $\alpha_\aaa$. 
A fractional brane $\FF_k$ is then characterized by a charge vector 
\be
{\rm ch}(\FF_k)= (\, r_k, \, p_{k}^{\,\aaa}, \, q_k \, )\, .
\ee
Here $r_k$ denotes  the D7 wrapping number,    $p_k^{\, \aaa}$ is
the D5 wrapping number around the 2-cycle $\alpha_\aaa$, and
$q_k$ counts the D3 charge. 
A fractional brane breaks half of the supersymmetries. Which 
half is preserved is determined by the central charge
\footnote{This expression for $\beta^\aaa$ is valid in the large volume limit only. }
 \ba
{\mathsf Z}(\FF_k) \is q_k +  p_{k}^{\, \aaa} \, {\beta}_\aaa- {\textstyle{1\over 2}}\, r_k \, \beta^\aaa \beta_\aaa\, ,\\[3mm]
&\beta^\aaa &\!\! 
=  \int_{\alpha_\aaa}\! (B + i J)\, .
\ee
A brane configuration assigns a multiplicity 
$n_k$ to each fractional brane $\FF_k$. For a supersymmetric arrangement, the central
charges of the fractional branes (times $n_k$)  all align.
 
 The $U(1)$ gauge
 symmetry associated with a fractional brane typically gets lifted via coupling to RR-forms. 
 A $U(1)$ factor only remains massless if it is associated
with a trivial 2-cycle with zero size.\footnote{Otherwise it
 would be associated with a non-zero FI-parameter, that via the D-term equations would dictate
 symmetry breaking expectations values for charged matter.} 
 In D-brane realizations of SM-like gauge theories,
this observation is of direct use in selecting hypercharge as the only massless $U(1)$
gauge symmetry. 
Inevitably, however, the hypercharge generator will be a linear
combination of the $U(1)$ generators. 
In other
words, hypercharge is associated with a linear combination of branes 
\ba
\label{sumf}
\FF_Y = \sumk\; \raisebox{.5mm}{$m_k $}\, \FF_k\, ,
\ea
with $m_k$ some set of (small) integers.  
The quantum numbers of $\FF_Y$ are those of a D5-brane wrapping a trivial 2-cycle within $X$,
with zero size, and it may
carry arbitrary D3 charge.\footnote{The $U(1)$ gauge symmetry associated with the D3-brane
decouples. In the model of \cite{BMMVW}, which is based on a single D3, adding one unit of
$D3$ charge amounts to a shifting each $m_k$ in (\ref{sumf}) by $n_k$.} 
A specific construction of an SM-like theory along these lines, complete with an explicit description
of its CY embedding, has recently been presented in~\cite{BMMVW}. 

As we have seen, brane configurations of this type are typically
metastable. In the present
case, however, the decay must
proceed via a collective tunneling process, in which the exact linear combination (\ref{sumf})
of fractional branes associated with hypercharge first forms a single D5-brane wrapping the
trivial cycle, which then unwraps itself along some 3-chain within the CY. 
Evidently, the decay probability needs to be small to be consistent with observation. 

For the specific construction of \cite{MH,BMMVW}, the SM was obtained via 
a collection of fractional branes with the quantum numbers of a single D3-brane.
The total configuration was argued 
to be supersymmetric, provided the local CY geometry can be appropriately 
adjusted.\footnote{This argument may need some further study, however. A possible point
of worry is that a trivial 2-cycle with finite area can not be supersymmetric. Note, however,
the `hypercharge brane' $\FF_Y$ that wraps the trivial cycle is just a virtual linear combination (\ref{sumf})
of the real (supersymmetric) fractional branes. Moreover, we assume that the
trivial 2-cycle has zero area, which may potentially restore its supersymmetry.
In general, however, we expect that the presence of a trivial 2-cycle introduces an interesting, perhaps subtle or possibly less subtle,  source of supersymmetry breaking that deserves some further exploration.}
After the hypercharge brane (\ref{sumf}) has unwrapped, the left-over brane configuration
is specified by the same basis of fractional branes, but with new multiplicities $n'_k = n_k-m_k$.
For the model of \cite{BMMVW} and with minimal $H_3$-flux ($K=0$ in (\ref{hflux})), it turns out that these new multiplicities $n'_k$ are either 0 or have the same sign
as the $n_k$. Thus, assuming that the initial collection is supersymmetric, we may conclude that
the final brane configuration is also supersymmetric. We thus expect that, for this specific model,
the decay probability is highly suppressed.

Another interesting feature of this scenario is the appearance of the stringy monopole 
solutions described in section 3.1. These monopoles carry one unit of magnetic
hypercharge, and their mass is typically of order the string scale. 
Evidently, the
origin of this new type of monopole is unrelated to the embedding of the SM gauge group inside
a Grand Unified gauge group. It is interesting to speculate, therefore, how small
the monopole mass can be in some suitable warped scenario, where the CY singularity
that supports the SM brane and the 3-chain $\Gamma$ are both located near the IR
region of a KS like geometry. The string scale at the bottom of the warped region can presumably 
be made as low as several TeV. The volume of the 3-chain $\Gamma$ needs to be
just a couple of times the unit volume in these string units, which is sufficient to
ensure that the D5-brane decay rate (\ref{finalrate}) is highly suppressed. A rough estimate of the
lowest possible monopole mass in this -- admittedly speculative -- 
class of scenarios is then of order 100 TeV or even
a bit less.\footnote{The existence of such light monopoles will
pose no problems for cosmology, provided they are not
produced during inflation
and the reheating temperature is low enough.}

For the purpose of self-education only, we end this section by writing the known Ansatz \cite{cho}
for a hypercharge monopole solution (with point source) after electro-weak
 symmetry breaking
\ba
\phi\!\is\! \rho(r) \xi \qquad \qquad \qquad \qquad \qquad \xi = \mbox{\small $\left(\!\! \begin{array}{c} e^{-i\varphi} \sin (\theta/2) \\ 
- \cos (\theta/2) \end{array}\!\! \right) $}\, ,
\\[3mm] 
B_\mu  \! \is \!   {\raisebox{-.5mm}{\small 1} \over\raisebox{.5mm}{$g'$}}(1-\cos \theta) \partial_\mu \varphi  \qquad \qquad \ \ A_\mu =  {\raisebox{-.5mm}{\small 1} \over\raisebox{.5mm}{$g$}} (f(r)-1) \,{\bf \hat{r}} \times \partial_\mu {\bf \hat r}\, .
\nonumber
\ea
Here $\rho(0)\!=\!0$ and approaches the constant Higgs vev at large $r$, and
$f(0)\!=\!1$ and goes to zero for large $r$.
Note that for this Ansatz  $A_\mu^3 \equiv  \xi^\dagger {\bf A_\mu} \xi = {\bf \hat r \cdot\! A_\mu} = 0$. Moreover, the whole $SU(2)$ part of the gauge field is smooth; the only singularity of the solution
is the point-like source of magnetic hypercharge at the origin.
In the unitary gauge, on the other hand, we have
\ba
\qquad \xi = \mbox{\small $\left(\!
\begin{array}{c} 0 \\ 
1 \end{array}\!\! \right) $}\, ,
\qquad \qquad \qquad \qquad
A^3_\mu =  {\raisebox{-.5mm}{\small 1} \over\raisebox{.5mm}{$g$}} (1-\cos \theta) \partial_\mu \varphi \, .
\ea
We see that the electro-weak symmetry breaking transmutes the magnetic hypercharge of the monopole
into a pure EM magnetic charge. The $A^3_\mu$ part of the magnetic point source 
is neutralized via the presence of a cloud of $W$-bosons 
\be
W^\pm_\mu =  {\raisebox{-.5mm}{\small 1} \over\raisebox{.5mm}{$g$}}\, e^{\pm i\varphi}(\sin \theta\, \partial_\mu \varphi \pm i \partial_\mu \theta) f(r)\, .
\ee
Our brane configuration of section 3.1 nicely regulates the point source of magnetic hypercharge,
and thereby renders the above solution into a regular monopole of finite mass.

\bigskip

\bigskip

\bigskip

\bigskip

\newsection{Conclusion}

We have studied some physical aspects of two related types of metastable brane 
arrangements:
anti-D3-branes on a Calabi-Yau compactification with flux, and D5-branes wrapping topologically trivial
2-cycles. We have found that both types of branes host regular, finite energy monopole 
excitations. Inside the core of the monopoles, the world-brane locally unwraps by traversing a
3-chain $\Gamma$ inside the compact CY. The metastable branes have a finite lifetime 
against completely unwrapping along $\Gamma$. Given the topological correspondence 
between the two, it is natural to think of the brane decay process as a condensation 
of monopoles. This interpretation is further supported by the fact that electric charges on the
metastable branes,  the end-points of fundamental open strings, become
manifestly confined after the brane has unwrapped itself. 

\bigskip
\bigskip

\noindent
{\bf \large Note Added}

Geometrically metastable configurations were recently also studied in \cite{ABSV} in the form of 
D5 and anti D5-branes wrapping homologous  2-spheres.  The large $N$ dual
was conjectured to be described by a smooth CY manifold, with the branes on the 2-spheres replaced
by 3-spheres with flux \cite{Vafaflux}. The models of \cite{ABSV} exhibit similar monopole configurations
as studied here. Before any geometric transition, they are wrapped D3-branes that connect
the metastable D5-branes. After one of the two geometric transitions, they look like the configuration
of section 2.1, with the tip of the 3-chain located at the ${\bf S}^3$ formed by the transition. After both 
transitions, the monopoles become D3-branes that wrap a closed 
3-cycle.  It is natural to further study the physics of our monopole configurations with the help of
the holographic dictionary of D-branes that undergo geometric transitions.

\bigskip

\bigskip

\noindent
{\large \bf Acknowledgments}

It is a pleasure to thank Oliver deWolfe, Jaume Gomis, Shamit Kachru,  Dave Morrison, Joe Polchinski, Cumrun Vafa and Martijn Wijnholt for helpful
discussion and comments. I also thank the KITP in Santa Barbara for hospitality and support while part of this
work was done. This work was
supported by the National Science Foundation under grant PHY-0243680. Any opinions,
findings, and conclusions or recommendations expressed in  this material are
those of the authors and do not necessarily reflect the views of the National Science
Foundation.


\renewcommand{\Large}{\large}

\end{document}